\documentclass[english,12pt]{article}
\usepackage{array}
\usepackage{graphicx}
\usepackage{amssymb}
\usepackage[english]{babel}
\usepackage{amsmath}
\usepackage{multirow}
\usepackage{prettyref}
\usepackage{babel}
\usepackage{units}
\usepackage[latin1]{inputenc}
\usepackage{amsfonts}
\usepackage{amssymb}
\usepackage{babel}
\usepackage{color}
\usepackage{cite}
\def\@fmsl@sh#1#2#3{\m@th\ooalign{$\hfil#1\mkern#2/\hfil$\crcr$#1#3$}}
 \def\eq#1\en{\begin{equation}#1\end{equation}}
\def\s[#1,#2]{[#1\stackrel{\star}{,}#2]}
\def\sx[#1,#2]{[#1\stackrel{\star_{x}}{,}#2]}

\newcommand{\nc}{\newcommand}
\nc{\beq}{\begin{equation}}
\nc{\eeq}{\end{equation}}
\nc{\beqa}{\begin{eqnarray}}
\nc{\eeqa}{\end{eqnarray}}

\def\bc{\begin{center}}
\def\ec{\end{center}}

\def\to{\rightarrow}

\def\gsim{\mathrel{\mathpalette\atversim>}}

\def\bc{\begin{center}}
\def\ec{\end{center}}

\def\gsim{\mathrel{\rlap{\lower4pt\hbox{\hskip1pt$\sim$}}

    \raise1pt\hbox{$>$}}}       %greater than or approx. symbol

\def\gsim{\mathrel{\rlap{\lower4pt\hbox{\hskip1pt$\sim$}}
    \raise1pt\hbox{$>$}}}       %greater than or approx. symbol

\begin{document}
\makeatletter
\def\fmslash{\@ifnextchar[{\fmsl@sh}{\fmsl@sh[0mu]}}
\def\fmsl@sh[#1]#2{%
  \mathchoice
    {\@fmsl@sh\displaystyle{#1}{#2}}%
    {\@fmsl@sh\textstyle{#1}{#2}}%
    {\@fmsl@sh\scriptstyle{#1}{#2}}%
    {\@fmsl@sh\scriptscriptstyle{#1}{#2}}}
\def\@fmsl@sh#1#2#3{\m@th\ooalign{$\hfil#1\mkern#2/\hfil$\crcr$#1#3$}}
\makeatother
%\baselineskip 24pt

%%%%%%%%%%%%%%%%%%%%%%%%%%%%%%%%%%%%%%%%%%%%%%%%%%%%%%%%%%%%%%%%%
%%%
%%%                      TITLE PAGE
%%%
%%%%%%%%%%%%%%%%%%%%%%%%%%%%%%%%%%%%%%%%%%%%%%%%%%%%%%%%%%%%%%%%%
\thispagestyle{empty}
\begin{titlepage}
\boldmath
\begin{center}
  \Large {\bf   Dark Matter in Quantum Gravity}
    \end{center}
\unboldmath
\vspace{0.2cm}
\begin{center}
{  {\large Xavier Calmet}\footnote{x.calmet@sussex.ac.uk}$^{a,b}$}{\large and}  
{  {\large Boris Latosh}\footnote{b.latosh@sussex.ac.uk}$^{a,c}$} 
 \end{center}
\begin{center}
$^a${\sl Department of Physics and Astronomy, 
University of Sussex, Brighton, BN1 9QH, United Kingdom
}\\
$^b${\sl PRISMA Cluster of Excellence and Mainz Institute for Theoretical Physics, Johannes Gutenberg University, 55099 Mainz, Germany }\\
$^c${\sl Dubna State University,
Universitetskaya str. 19, Dubna 141982, Russia}
\end{center}
\vspace{5cm}
\begin{abstract}
\noindent
We show that quantum gravity, whatever its ultra-violet completion might be, could account for dark matter. Indeed, besides the massless gravitational field recently observed in the form of gravitational waves, the spectrum of quantum gravity contains two massive fields respectively of spin 2 and spin 0. If these fields are long-lived, they could easily account for dark matter. In that case, dark matter would be very light and only gravitationally coupled to the standard model particles. 
\end{abstract}  
\vspace{5cm}
%Essay written for the Gravity Research Foundation 2018 Awards for Essays on Gravitation.
\end{titlepage}

%\pacs{}

%%%%%%%%%%%%%%%%%%%%%%%%%%%%%%%%%%%%%%%%%%%%%%%%%%%%%%%%%%%%%%%%
%%%
%%%                     INTRODUCTION
%%%
%%%%%%%%%%%%%%%%%%%%%%%%%%%%%%%%%%%%%%%%%%%%%%%%%%%%%%%%%%%%%%%%

\newpage

While finding a unified theory of quantum field theory and general relativity remains an elusive goal, much progress has been done recently in quantum gravity using effective field theory methods \cite{Donoghue:1994dn,Donoghue:1993eb,BjerrumBohr:2002kt,Calmet:2013hfa,Donoghue:2014yha,Calmet:2015dpa,Alexeyev:2017scq,Calmet:2008tn,Calmet:2017qqa,Calmet:2016sba,Calmet:2017rxl,Calmet:2014gya,Calmet:2015pea,Calmet:2017omb,Calmet:2018qwg}. This approach enables one to perform model independent calculations in quantum gravity. The only restriction is that only physical processes taking place at energy scales below the Planck mass can be considered. This restriction is, however, not very constraining as this is the case for all practical purposes in particle physics, astrophysics and cosmology.

In this paper, we show that quantum gravity could provide a solution to the long standing problem of dark matter. There are overwhelming astrophysical and cosmological evidences that visible matter only constitutes a small fraction of the total matter of our universe and that most of it is a new form of non-relativistic dark matter which cannot be accounted for by the standard model of particle physics. Gravity could account for dark matter in two forms. The first gravitational dark matter candidates are primordial black holes, see e.g. \cite{Carr:2016drx} for a recent review. They have been investigated for many years, and although the mass range for such objects to account for dark matter has shrunk quite a bit, they remain a viable option for dark matter, in particular Planckian mass black hole remnants are good dark matter candidates. Here we discuss a second class of candidates within the realm on quantum gravity. Recent work in quantum gravity has established in a model independent way that the spectrum of quantum gravity involves, beyond the massless gravitational field already observed in the form of gravitational waves, two new massive fields \cite{Calmet:2014gya}. Their properties can be derived from the effective action for quantum gravity. We will show here that these new fields are ideal dark matter candidates.

Deriving an effective action for quantum gravity requires starting from general relativity and integrating out fluctuations of the graviton. Doing so, we obtain a classical effective action given at second order in curvature by
\begin{eqnarray}\label{action1}
S &=& \int d^4x \, \sqrt{-g} \left[ \left( \frac{1}{2}  M^2 + \xi H^\dagger H \right)  \mathcal{R}- \Lambda_C^4 + c_1 \mathcal{R}^2 + c_2 \mathcal{R}_{\mu\nu}\mathcal{R}^{\mu\nu}+ c_4   \Box \mathcal{R}  \right . \nonumber \\
&& \left . - b_1 \mathcal{R} \log \frac{\Box}{\mu^2_1}\mathcal{R} - b_2 \mathcal{R}_{\mu\nu}  \log \frac{\Box}{\mu^2_2}\mathcal{R}^{\mu\nu}  
- b_3 \mathcal{R}_{\mu\nu\rho\sigma}  \log \frac{\Box}{\mu^2_3}\mathcal{R}^{\mu\nu\rho\sigma} 
+ \mathcal{L}_{SM}+ \mathcal{O}(M_\star^{-2})   \right],
\end{eqnarray}
where $\mathcal{R}$, $\mathcal{R}_{\mu\nu}$ and $\mathcal{R}_{\mu\nu\rho\sigma}$ are respectively the Ricci scalar, the Ricci tensor and the Riemann tensor.   The cosmological constant is denoted by $\Lambda_C$. The scales $\mu_i$ are renormalization scales which in principle could be different, we shall however take $\mu_i=\mu$. The Lagrangian $L_{SM}$ contains all of the matter we know of and $M_\star$ is the energy scale up to which we can trust the effective field theory. The term $\Box \mathcal{R}$ is a total derivative and thus does not contribute to the equation of motions. 

Remarkably, the values of the parameters $b_i$ are calculable from first principles and are model independent predictions of quantum gravity, see e.g. \cite{Birrell:1982ix} and references therein. They are related to the number of fields that have been integrated out. The non-renormalizability  of the effective action is reflected in the fact that we cannot predict the coefficients $c_i$ which, in this framework, have to be measured in experiments or observations. There will be new $c_i$ appearing at every order in the curvature expansion performed when deriving this effective action and we thus would have to measure an infinite number of parameters. Despite this fact, the effective theory leads to falsifiable predictions as the coefficients $b_i$ of  non-local operators are, as explained previously, calculable. 

In \cite{Calmet:2017rxl,Calmet:2018qwg}, it was shown how to identify the new degrees of freedom by finding the poles of the Green's function obtained by varying the linearized version of the action given in Eq.(\ref{action1}) with respect to the metric. Besides the usual massless pole, one finds two pair of complex poles. The complex pole for the massive spin-2 object is given by
\begin{align}
m_2^2&=\frac{2}{ (b_2+ 4 b_3) \kappa^2 W\left(-\frac{2 \exp\frac{-c_2}{(b_2+ 4 b_3)}}{ (b_2+ 4 b_3) \kappa^2 \mu^2}\right)},
\end{align}
while that of the massive spin-0 reads
\begin{align}
m_0^2&=\frac{-1}{ (3 b_1+b_2+ b_3) \kappa^2 W\left(\frac{ \exp\frac{-3 c_1-c_2}{(3 b_1+b_2+ b_3)}}{ (3 b_1+b_2+ b_3) \kappa^2 \mu^2}\right)},
\end{align}
where $W(x)$ is the Lambert function and $\kappa^2=32 \pi G$, $G$ is Newton's constant. The $b_i$ for the graviton are known: $b_1=430/(11520 \pi^2)$, $b_2=-1444/(11520 \pi^2)$ and $b_3=434/(11520 \pi^2)$. The $b_i$ are thus small and unless the $c_i$ are large, the masses $m_2$ and $m_0$ will be close to the Planck mass $M_P$ and the corresponding fields will decay almost instantaneously \cite{Calmet:2014gya}. As we are interested in the case where the new fields are light, it is useful to consider the limit where the $c_i$ (or one of them at least) are large and $b_i\ll c_i$.  In that case we can rewrite the masses as
\begin{align}
m_2^2&= -\frac{2}{ \kappa^2 c_2}  - i \pi \frac{2}{ \kappa^2 c_2^2}  (b_2+ 4 b_3),
\end{align}
so we need to pick $c_2<0$ and
\begin{align}
m_0^2&= \frac{1}{ \kappa^2 (3 c_1+c_2)}  - i \pi \frac{1}{ \kappa^2 (3 c_1+c_2)^2}  (3 b_1+b_2+ b_3),
\end{align}
where we assumed that the renormalization scale $\mu\sim 1/\kappa$, i.e. we assume that the effective field theory is valid up to the reduced Planck scale. As done in \cite{Calmet:2014gya}, we can identify the mass and width of the respective field using $m^2_{i}=(M_{i}-i \Gamma_{i}/2)^2$. Note that the complex conjugate solutions $m_2^\star$ and $m_0^\star$ which lead to a positive sign between the mass and the width in the propagator can be eliminated by a proper choice of the contour integral, i.e. of boundary conditions\cite{Calmet:2017omb}, in full analogy with the usual $i\epsilon$ procedure which enables one to select the causal behavior of the Green's function.

 We can now express the width in terms of the mass of the field. For the massive spin-2 field $k$, we find
 \begin{align}
M_2 &=\sqrt{\frac{2}{c_2}} \frac{M_{P}}{2}, \\
\Gamma_2 &\approx   \frac{(b_2+ 4 b_3) \pi}{\sqrt{2 c_2^3}}M_{P}=
\frac{73 M_2^3}{360 \pi \sqrt{2} M_P^2},
\end{align}
and for the massive spin-0 field $\sigma$, one has
 \begin{align}
M_0 &\approx \sqrt{\frac{1}{(3 c_1+c_2) \kappa^2}}=\sqrt{\frac{1}{(3 c_1+c_2)}} \frac{M_P}{2}, \\
\Gamma_0 &\approx   \frac{(3 b_1+b_2+ b_3) \pi}{2 \sqrt{ (3 c_1+c_2)^3}}M_P=\frac{7 M_0^3}{72 \pi  M_P^2},
\end{align}
where $M_{P}=2.435\times 10^{18}$ GeV is the reduced Planck mass. The widths $\Gamma_0$ and $\Gamma_2$ are the gravitational widths for the decay of the massive spin-2 and spin-0 classical modes into the classical graviton. 

To obtain the total width, we need to include the decay modes into particles of the standard model. The coupling of the two states to the standard model Lagrangian has been worked out in \cite{Calmet:2018qwg}.  One has
\begin{align}
S=\int d^4 x \left[\left (- \frac{1}{2} h_{\mu\nu} \Box h^{\mu\nu}
 +\frac{1}{2} h_{\mu}^{\ \mu} \Box h_{\nu}^{\ \nu}  -h^{\mu\nu} \partial_\mu \partial_\nu h_{\alpha}^{\ \alpha}+ h^{\mu\nu} \partial_\rho \partial_\nu h^{\rho}_{\ \mu}\right) \right. \\ 
\left. \nonumber + \left ( -\frac{1}{2} k_{\mu\nu} \Box k^{\mu\nu}
 +\frac{1}{2} k_{\mu}^{\ \mu} \Box k_{\nu}^{\ \nu}  -k^{\mu\nu} \partial_\mu \partial_\nu k_{\alpha}^{\ \alpha}+ k^{\mu\nu} \partial_\rho \partial_\nu k^{\rho}_{\ \mu}
 \right.  \right.  \\ \left. \left. \nonumber
 -\frac{M_2^2}{2} \left (k_{\mu\nu}k^{\mu\nu} - k_{\alpha}^{\ \alpha} k_{\beta}^{\ \beta} \right )
 \right)  \right.
  \\
  \nonumber
 \left. + \frac{1}{2} \partial_\mu \sigma  \partial^\mu \sigma
  - \frac{M_0^2}{2} \sigma^2 - \sqrt{8 \pi G_N} (h_{\mu\nu}-k_{\mu\nu}+\frac{1}{\sqrt{3}} \sigma \eta_{\mu\nu})T^{\mu\nu} 
  \right ].
\end{align}
We thus see that besides decaying gravitationally, the massive spin-2 and spin-0 fields can decay to standard model particles. It is straightforward to calculate the decay widths of the new massive modes into standard model particles using the results of \cite{Han:1998sg}.

The decay width of the scalar mode $\sigma$ into massive vectors fields $V$, such as the W and Z bosons, is given by
\begin{align}
\Gamma(\sigma \to V V)= \delta \frac{M_0^3 }{48 \pi M_{P}^2} \left (1- 4 r_V \right)^{1/2} \left(1 - 4 r_V+12 r_V^2 \right),
\end{align}
where $\delta=1/2$ for identical particles and $r_V=(m_V/M_0)^2$. The decay width of $\sigma$ into fermions is given by
\begin{align}
\Gamma(\sigma \to \bar{f} f)= \frac{m_f^2 M_0 N_c}{24 \pi M_{P}^2} \left (1- 4 r_f \right)^{1/2}
\left (1- 2 r_f \right)
\end{align}
with $r_f=(m_f/M_0)^2$ and $N_C=3$ if the fermions are quarks. While $\sigma$ couples to the trace of the energy-momentum tensor of the standard model and it thus does not couple to massless gauge bosons at tree level, it will couple to the photon and the gluons at one loop. In particular the decay width into two photons is given by \cite{Goldberger:2008zz,Cembranos:2008gj}
\begin{align}
\Gamma(\sigma \to \gamma \gamma)= \frac{\alpha_{EM}^2 M_0^3 N_c}{768 \pi^3 M_{P}^2} |c_{EM}|^2,
\end{align}
where $\alpha_{EM}$=1/137 and $c_{EM}$=11/3 if $\phi$ is lighter than all the fermions of the standard model. The decay width of $\sigma$ into a pair of Higgs bosons is given by
\begin{align}
\Gamma(\sigma \to h h)= \frac{M_0^3}{48 \pi M_{P}^2} \left (1- 4 r_h \right)^{1/2}
\left (1+ 2 r_h \right)^2,
\end{align}
where $r_h=(m_h/M_0)^2$.

It is also straightforward to calculate the partial decay widths of the spin-2 object $k$. Its partial width to massless vector fields is given by
\begin{align}
\Gamma(k \to V V) = N \frac{M_2^3}{80 \pi M_P^2},
\end{align}
where $N$=1 for photons and $N=8$ for gluons. In the case of massive massive vector fields, one has
\begin{align}
\Gamma(k \to V V) = \delta  \frac{M_2^3}{40 \pi M_P^2} \sqrt{1-4 r_V} \left (\frac{13}{12}+\frac{14}{3} r_V + \frac{4}{13} r_V^2 \right ),
\end{align}
where $\delta=1/2$ for identical particles, $r_V=m^2_V/M_2^2$. For the decay to fermions, we find 
\begin{align}
\Gamma(k \to \bar f f) = N_C  \frac{M_2^3}{160 \pi M_P^2}
  \left ( 1- 4 r_f \right )^{3/2} \left (1+\frac{8}{3} r_f \right ),
\end{align}
where $r_f=m^2_f/M_2^2$ and, as previously, $N_C=3$ if the fermions are quarks. In the case of a decay to the Higgs boson, the partial decay width is given by
\begin{align}
\Gamma(k \to h h) =   \frac{M_2^3}{430 \pi M_P^2}
  \left ( 1- 4 r_h \right )^{5/2},
\end{align}
where $r_h=m^2_h/M_2^2$.

If the massive spin-0 and spin-2 fields are components of the dark matter content of the universe nowadays, their masses have to be such that none of these partial decay widths should enable these fields to decay faster than the current age of the universe. 
From the requirement that the lifetime of the spin-0 $\sigma$ is longer than current age of the universe, we can thus get a bound on $c_2$ using the gravitational decay width. We find
\begin{align}
\tau=1/\Gamma= 7.2 \times 10^{-17} \sqrt{c_2^3}  \ \mbox{GeV}^{-1}>13.77 \times 10^9 \mbox{y}
 \end{align}
and thus $c_2>4.4 \times 10^{38}$. The same reasoning leads to a similar bound on $3 c_1+c_2$. We can then deduce a maximal mass for the dark matter candidate, $M_0<0.16$ GeV. Note that E\"ot-Wash \cite{Hoyle:2004cw}
 implies $c_2<10^{61}$, we thus have a bound $4.4 \times 10^{38}<c_2<10^{61}$ and $1\times 10^{-12} \ \mbox{GeV}<M_0<0.16$ GeV. Again a similar bound applies to the combination $3c_1+c_2$ and thus to $M_2$. Clearly such light dark matter candidates could not decay to the massive gauge bosons of the standard model, its charged leptons such as the electron or the quarks. They could however decay to gluons (during the deconfinement phase of the early Universe), photons and potentially neutrinos. The decay to photons might be of astrophysical relevance and could be observable by gamma-ray experiments. Note, however, that decay widths of the dark matter candidates to photons are smaller than the respective gravitational ones. It is also worth mentioning that the decay to neutrinos can be as rapid as the gravitational modes if again neutrino masses are low enough.

While we have established that quantum gravity provides two new candidates for dark matter, it remains to investigate their production mechanism. Thermal production is a possibility, but we would have to consider all higher order operators as we would need to consider temperatures larger than the Planck mass $T\ge M_P$ since these objects are gravitationally coupled to all matter fields. Also we may not want to involve temperatures above the inflation scale which we know is at most $10^{14}$ GeV. The weakness of the Planck-suppressed coupling hints at the possibility of out-of-equilibrium thermal production as argued in \cite{Babichev:2016hir}. However, the mass range allowed for the dark matter particles within that framework is given by TeV$<m_{DM}< 10^{11}$ GeV \cite{Babichev:2016hir} and it is not compatible with our ranges for the masses of our candidates. The fact that our dark matter candidates are light points towards the vacuum misalignment mechanism, see e.g. \cite{Nelson:2011sf}. Indeed, in an expanding universe both $\sigma$ and $k$ have an effective potential in which they oscillate. The amount of dark matter produced by this mechanism becomes simply a randomly chosen initial condition for the value of the field in our patch of the universe. In \cite{Arias:2012az}, it was shown that the vacuum misalignment mechanism leads to the correct dark matter abundance $\rho_{DM}=1.17$ \mbox{keV}/{\mbox{cm}$^3$ if the dark matter field takes large values in the early universe. For example, a dark matter field with a mass in the eV region would need to take values of the order of $10^{11}$ GeV to account for all of the dark matter in today's universe \cite{Arias:2012az}.

In summary, we have shown that gravity, when quantized, provides new dark matter candidates. As these fields must live long enough to still be around in today's universe their masses must be light otherwise they would have decayed long ago. It is quite possible that gravity can account for all of dark matter in the form of primordial black holes and the new fields discussed in this paper without the need for new physics.

{\it Acknowledgments:}
The work of XC is supported in part  by the Science and Technology Facilities Council (grant number ST/P000819/1). XC is very grateful to MITP for their generous hospitality during the academic year 2017/2018. 

%%%%%%%%%%%%%%%%%%%%%%%%%%%%%%%%%%%%%%%%%%%%%%%%%%%%%%%%%%%%%%%%%
%%%
%%%                     BIBLIOGRAPHY
%%%
%%%%%%%%%%%%%%%%%%%%%%%%%%%%%%%%%%%%%%%%%%%%%%%%%%%%%%%%%%%%%%%%%

\bigskip{}


\baselineskip=1.6pt 

\begin{thebibliography}{10}

  %\cite{Donoghue:1994dn}
\bibitem{Donoghue:1994dn} 
  J.~F.~Donoghue,
  %``General relativity as an effective field theory: The leading quantum corrections,''
  Phys.\ Rev.\ D {\bf 50}, 3874 (1994)
  doi:10.1103/PhysRevD.50.3874
  [gr-qc/9405057].
  %%CITATION = doi:10.1103/PhysRevD.50.3874;%%
  %603 citations counted in INSPIRE as of 01 Dec 2017
  
  %\cite{Donoghue:1993eb}
\bibitem{Donoghue:1993eb} 
  J.~F.~Donoghue,
  %``Leading quantum correction to the Newtonian potential,''
  Phys.\ Rev.\ Lett.\  {\bf 72}, 2996 (1994)
  doi:10.1103/PhysRevLett.72.2996
  [gr-qc/9310024].
  %%CITATION = doi:10.1103/PhysRevLett.72.2996;%%
  %321 citations counted in INSPIRE as of 01 Dec 2017

%\cite{BjerrumBohr:2002kt}
\bibitem{BjerrumBohr:2002kt} 
  N.~E.~J.~Bjerrum-Bohr, J.~F.~Donoghue and B.~R.~Holstein,
  %``Quantum gravitational corrections to the nonrelativistic scattering potential of two masses,''
  Phys.\ Rev.\ D {\bf 67}, 084033 (2003)
  Erratum: [Phys.\ Rev.\ D {\bf 71}, 069903 (2005)]
  doi:10.1103/PhysRevD.71.069903, 10.1103/PhysRevD.67.084033
  [hep-th/0211072].
  %%CITATION = doi:10.1103/PhysRevD.71.069903, 10.1103/PhysRevD.67.084033;%%
  %191 citations counted in INSPIRE as of 05 Dec 2017
  

%\cite{Calmet:2013hfa}
\bibitem{Calmet:2013hfa} 
  X.~Calmet,
  %``Effective theory for quantum gravity,''
  Int.\ J.\ Mod.\ Phys.\ D {\bf 22}, 1342014 (2013)
  doi:10.1142/S0218271813420145
  [arXiv:1308.6155 [gr-qc]].
  %%CITATION = doi:10.1142/S0218271813420145;%%
  %8 citations counted in INSPIRE as of 01 Dec 2017
  
  %\cite{Donoghue:2014yha}
\bibitem{Donoghue:2014yha} 
  J.~F.~Donoghue and B.~K.~El-Menoufi,
  %``Nonlocal quantum effects in cosmology: Quantum memory, nonlocal FLRW equations, and singularity avoidance,''
  Phys.\ Rev.\ D {\bf 89}, no. 10, 104062 (2014)
  doi:10.1103/PhysRevD.89.104062
  [arXiv:1402.3252 [gr-qc]].
  %%CITATION = doi:10.1103/PhysRevD.89.104062;%%
  %36 citations counted in INSPIRE as of 04 Dec 2017
  
%\cite{Calmet:2015dpa}
\bibitem{Calmet:2015dpa} 
  X.~Calmet, D.~Croon and C.~Fritz,
  %``Non-locality in Quantum Field Theory due to General Relativity,''
  Eur.\ Phys.\ J.\ C {\bf 75}, no. 12, 605 (2015)
  doi:10.1140/epjc/s10052-015-3838-2
  [arXiv:1505.04517 [hep-th]].
  %%CITATION = doi:10.1140/epjc/s10052-015-3838-2;%%
  %12 citations counted in INSPIRE as of 01 Dec 2017
  
%\cite{Alexeyev:2017scq}
\bibitem{Alexeyev:2017scq} 
  S.~O.~Alexeyev, X.~Calmet and B.~N.~Latosh,
  %``Gravity induced non-local effects in the standard model,''
  Phys.\ Lett.\ B {\bf 776}, 111 (2018)
  doi:10.1016/j.physletb.2017.11.028
  [arXiv:1711.06085 [hep-th]].
  %%CITATION = doi:10.1016/j.physletb.2017.11.028;%%
  %1 citations counted in INSPIRE as of 01 Dec 2017
  
  
  %\cite{Calmet:2008tn}
\bibitem{Calmet:2008tn} 
  X.~Calmet, S.~D.~H.~Hsu and D.~Reeb,
  %``Quantum gravity at a TeV and the renormalization of Newton's constant,''
  Phys.\ Rev.\ D {\bf 77}, 125015 (2008)
  doi:10.1103/PhysRevD.77.125015
  [arXiv:0803.1836 [hep-th]].
  %%CITATION = doi:10.1103/PhysRevD.77.125015;%%
  %70 citations counted in INSPIRE as of 01 Dec 2017
  
  
%\cite{Calmet:2017qqa}
\bibitem{Calmet:2017qqa} 
  X.~Calmet and B.~K.~El-Menoufi,
  %``Quantum Corrections to Schwarzschild Black Hole,''
  Eur.\ Phys.\ J.\ C {\bf 77}, no. 4, 243 (2017)
  doi:10.1140/epjc/s10052-017-4802-0
  [arXiv:1704.00261 [hep-th]].
  %%CITATION = doi:10.1140/epjc/s10052-017-4802-0;%%
  %5 citations counted in INSPIRE as of 01 Dec 2017
  
  
  
  %\cite{Calmet:2016sba}
\bibitem{Calmet:2016sba} 
  X.~Calmet, I.~Kuntz and S.~Mohapatra,
  %``Gravitational Waves in Effective Quantum Gravity,''
  Eur.\ Phys.\ J.\ C {\bf 76}, no. 8, 425 (2016)
  doi:10.1140/epjc/s10052-016-4265-8
  [arXiv:1607.02773 [hep-th]].
  %%CITATION = doi:10.1140/epjc/s10052-016-4265-8;%%
  %6 citations counted in INSPIRE as of 01 Dec 2017
  

%\cite{Calmet:2017rxl}
\bibitem{Calmet:2017rxl} 
  X.~Calmet, S.~Capozziello and D.~Pryer,
  %``Gravitational Effective Action at Second Order in Curvature and Gravitational Waves,''
  Eur.\ Phys.\ J.\ C {\bf 77}, no. 9, 589 (2017)
  doi:10.1140/epjc/s10052-017-5172-3
  [arXiv:1708.08253 [hep-th]].
  %%CITATION = doi:10.1140/epjc/s10052-017-5172-3;%%
  



%\cite{Calmet:2014gya}
\bibitem{Calmet:2014gya} 
  X.~Calmet,
  %``The Lightest of Black Holes,''
  Mod.\ Phys.\ Lett.\ A {\bf 29}, no. 38, 1450204 (2014)
  doi:10.1142/S0217732314502046
  [arXiv:1410.2807 [hep-th]].
  %%CITATION = doi:10.1142/S0217732314502046;%%
  %23 citations counted in INSPIRE as of 15 Mar 2018
  
%\cite{Calmet:2015pea}
\bibitem{Calmet:2015pea} 
  X.~Calmet and R.~Casadio,
  %``The horizon of the lightest black hole,''
  Eur.\ Phys.\ J.\ C {\bf 75}, no. 9, 445 (2015)
  doi:10.1140/epjc/s10052-015-3668-2
  [arXiv:1509.02055 [hep-th]].
  %%CITATION = doi:10.1140/epjc/s10052-015-3668-2;%%
  %15 citations counted in INSPIRE as of 15 Mar 2018
  
  %\cite{Calmet:2017omb}
\bibitem{Calmet:2017omb} 
  X.~Calmet, R.~Casadio, A.~Y.~Kamenshchik and O.~V.~Teryaev,
  %``Graviton propagator, renormalization scale and black-hole like states,''
  Phys.\ Lett.\ B {\bf 774}, 332 (2017)
  doi:10.1016/j.physletb.2017.09.080
  [arXiv:1708.01485 [hep-th]].
  %%CITATION = doi:10.1016/j.physletb.2017.09.080;%%
  %4 citations counted in INSPIRE as of 15 Mar 2018
  
%\cite{Calmet:2018qwg}
\bibitem{Calmet:2018qwg} 
  X.~Calmet and B.~Latosh,
  %``Three Waves for Quantum Gravity,''
   Eur.\ Phys.\ J.\ C {\bf 78}, 205 (2018)
  [arXiv:1801.04698 [hep-th]].
  %%CITATION = ARXIV:1801.04698;%%
  
  
  
  %\cite{Carr:2016drx}
\bibitem{Carr:2016drx} 
  B.~Carr, F.~Kuhnel and M.~Sandstad,
  %``Primordial Black Holes as Dark Matter,''
  Phys.\ Rev.\ D {\bf 94}, no. 8, 083504 (2016)
  doi:10.1103/PhysRevD.94.083504
  [arXiv:1607.06077 [astro-ph.CO]].
  %%CITATION = doi:10.1103/PhysRevD.94.083504;%%
  %161 citations counted in INSPIRE as of 15 Mar 2018
  
  
  
  %\cite{Birrell:1982ix}
\bibitem{Birrell:1982ix} 
  N.~D.~Birrell and P.~C.~W.~Davies,
``Quantum Fields in Curved Space,'' Cambridge University Press, Cambridge, 1982, 
  doi:10.1017/CBO9780511622632
  %%CITATION = doi:10.1017/CBO9780511622632;%%
  %1445 citations counted in INSPIRE as of 10 Jul 2017

%\cite{Hoyle:2004cw}
\bibitem{Hoyle:2004cw} 
  C.~D.~Hoyle, D.~J.~Kapner, B.~R.~Heckel, E.~G.~Adelberger, J.~H.~Gundlach, U.~Schmidt and H.~E.~Swanson,
  %``Sub-millimeter tests of the gravitational inverse-square law,''
  Phys.\ Rev.\ D {\bf 70}, 042004 (2004)
  doi:10.1103/PhysRevD.70.042004
  [hep-ph/0405262].
  %%CITATION = doi:10.1103/PhysRevD.70.042004;%%
  %322 citations counted in INSPIRE as of 11 Nov 2017

  %\cite{Han:1998sg}
\bibitem{Han:1998sg}
  T.~Han, J.~D.~Lykken and R.~J.~Zhang,
  %``On Kaluza-Klein states from large extra dimensions,''
  Phys.\ Rev.\ D {\bf 59} (1999) 105006
  doi:10.1103/PhysRevD.59.105006
  [hep-ph/9811350].
  %%CITATION = doi:10.1103/PhysRevD.59.105006;%%
  %957 citations counted in INSPIRE as of 16 Mar 2018


%\cite{Goldberger:2008zz}
\bibitem{Goldberger:2008zz} 
  W.~D.~Goldberger, B.~Grinstein and W.~Skiba,
  %``Distinguishing the Higgs boson from the dilaton at the Large Hadron Collider,''
  Phys.\ Rev.\ Lett.\  {\bf 100}, 111802 (2008)
  doi:10.1103/PhysRevLett.100.111802
  [arXiv:0708.1463 [hep-ph]].
  %%CITATION = doi:10.1103/PhysRevLett.100.111802;%%
  %267 citations counted in INSPIRE as of 19 Mar 2018
  
    %\cite{Cembranos:2008gj}
\bibitem{Cembranos:2008gj} 
  J.~A.~R.~Cembranos,
  %``Dark Matter from R2-gravity,''
  Phys.\ Rev.\ Lett.\  {\bf 102}, 141301 (2009)
  doi:10.1103/PhysRevLett.102.141301
  [arXiv:0809.1653 [hep-ph]].
  %%CITATION = doi:10.1103/PhysRevLett.102.141301;%%
  %98 citations counted in INSPIRE as of 04 Mar 2018
  
%\cite{Babichev:2016hir}
\bibitem{Babichev:2016hir} 
  E.~Babichev, L.~Marzola, M.~Raidal, A.~Schmidt-May, F.~Urban, H.~Veermäe and M.~von Strauss,
  %``Bigravitational origin of dark matter,''
  Phys.\ Rev.\ D {\bf 94}, no. 8, 084055 (2016)
  doi:10.1103/PhysRevD.94.084055
  [arXiv:1604.08564 [hep-ph]].
  %%CITATION = doi:10.1103/PhysRevD.94.084055;%%
  %20 citations counted in INSPIRE as of 19 Mar 2018
  
    %\cite{Nelson:2011sf}
\bibitem{Nelson:2011sf} 
  A.~E.~Nelson and J.~Scholtz,
  %``Dark Light, Dark Matter and the Misalignment Mechanism,''
  Phys.\ Rev.\ D {\bf 84}, 103501 (2011)
  doi:10.1103/PhysRevD.84.103501
  [arXiv:1105.2812 [hep-ph]].
  %%CITATION = doi:10.1103/PhysRevD.84.103501;%%
  %88 citations counted in INSPIRE as of 04 Mar 2018
  
   %\cite{Arias:2012az}
\bibitem{Arias:2012az} 
  P.~Arias, D.~Cadamuro, M.~Goodsell, J.~Jaeckel, J.~Redondo and A.~Ringwald,
  %``WISPy Cold Dark Matter,''
  JCAP {\bf 1206}, 013 (2012)
  doi:10.1088/1475-7516/2012/06/013
  [arXiv:1201.5902 [hep-ph]].
  %%CITATION = doi:10.1088/1475-7516/2012/06/013;%%
  %242 citations counted in INSPIRE as of 22 May 2018 
 


  
\end{thebibliography}
\end{document}